%Paper: hep-ph/9308240
%From: YUAN@MSUPA.PA.MSU.EDU
%Date: Fri, 6 Aug 1993 17:04:45 -0400 (EDT)

\input phyzzx
\overfullrule=0pt
\hsize=5in
\vsize=8in
\hoffset=0.5in
\voffset=0.5in
%========================DEFINITIONS===================================
\def\ETslash{\not{\hbox{\kern-4pt $E_T$}}}

\def\ra{\rightarrow}

\def\tbW {$t$-$b$-$W$ }

\def\width{ \Gamma( t \ra b W^+) }
\def\refmark#1{~[#1]}
%%%%
\def\jourtpol#1&#2{{\it #1}{\bf #2}}
\def\journal#1&#2(#3)#4{
{\unskip,~\it #1\unskip~\bf\ignorespaces #2\unskip~\rm (19#3) #4}}

\nopubblock
\line{\hfil MSUHEP-93/10 }
\line{\hfil July 1993}

\titlepage
\title{ Top Quark Physics at Hadron Colliders
\foot{Talk presented at the Workshop on ``Particle Physics at the Fermi
Scale'', Beijing, China, May  27 -- June 4, 1993.}
}
\author{ C.--P. Yuan}
\address{
Department of Physics and Astronomy \break
Michigan State University  \break
East Lansing, MI 48824}

\singlespace
\abstract{
I briefly report on what we can learn about the top quark
at hadron colliders. }

\endpage
%=========================Body of paper=======================
\normalspace

The top quark has been found to be heavier than 45 GeV from
SLAC and LEP experiments
and 108 (103) GeV from CDF (D0) data.
The first limit is model independent while the second limit is for the
Standard Model (SM) top quark.
Upper bounds on $m_t$ can be obtained from examining the
radiative corrections to low energy observables, such as the $\rho$
parameter which is proportional to $m_t^2$ at the one loop level
and thus sensitive to $m_t$.
The consistency of all the low energy experimental data
requires $m_t$ to be less than about
200 GeV.

\REF\toppol{
G.L.~Kane, G.A.~Ladinsky and C.-P.~Yuan \journal Phys.~Rev.&D45 (92) 124.}
\REF\decay{
I.I.Y. Bigi, Yu L. Dokshitzer, V.A. Khoze, J.H. Kuhn, and P. Zerwas
\journal Phys.~Lett.&B181 (86) 157;
L.H. Orr and J.L. Rosner, {\it Phys.~Lett.} {\bf B246} (1990) 221;
{\bf 248B} (1990) 474(E).}

Since the top quark is heavy, of the
same order of magnitude as the $W$--boson mass, any physical observable
related to the top quark may be sensitive to new physics.
The top quarks will therefore allow many new tests of the SM and new probes of
physics at the 100 GeV scale\refmark{\toppol}.
The most important consequence of a heavy top quark
is that to a good approximation it decays as a free quark since its
lifetime is short and it does not have time to
bind with light quarks before it decays\refmark{\decay}.
 Thus we can use the polarization
properties of the top quark
as additional observables to test the SM and
to probe new physics.
 Furthermore, because the heavy top quark
has the weak two--body decay $t \ra b W^+$, it will
analyze its own polarization.

In this brief report, I will discuss what we can learn about
the top quark at hadron colliders. First, I will discuss the production
mechanism for top quarks, then I will
discuss the decay modes and branching ratios
for top quarks. I will also report on how well the mass and
width of the top quark can be measured, and
how to detect CP violation effects from studying top quarks.

\REF\nloqcd{
P. Nason, S. Dawson and R.K. Ellis, \jourtpol
Nucl.~Phys. &{B303}, 607 (1988); {\bf B327}, 49 (1989);
W. Beenakker, H. Kuijf, W.L. van Neerven and J. Smith,
\jourtpol Phys.~Rev. &{D40}, 54 (1989);
R. Meng, G.A. Schuler, J. Smith and W.L. van Neerven,
\jourtpol Nucl.~Phys. &{B339}, 325 (1990).}

\REF\wgone{C.--P. Yuan, \jourtpol Phys.~Rev. &{D41}, 42 (1990);
D. Carlson and C.--P. Yuan, {\it Phys. Lett.} {\bf B306} (1993) 386.}
\REF\wgtwo{S. Dawson, \jourtpol Nucl.~Phys. &{B249}, 42 (1985);
S. Dawson and S. S. D. Willenbrock, \jourtpol Nucl.~Phys. &{B284}, 449 (1987);
S. S. D. Willenbrock and D. A. Dicus, \jourtpol Phys.~Rev. &{D34}, 155 (1986);
F. Anselmo, B. van Eijk and G. Bordes, \jourtpol Phys.~Rev. &{D45}, 2312
(1992);
T. Moers, R. Priem, D. Rein and H. Reithler in Proceedings of Large
Hadron Collider Workshop, preprint CERN 90-10, 1990;
R.K. Ellis and S. Parke, \jourtpol Phys.~Rev. &{D46}, 3785 (1992).}

At the Tevatron, the dominant production mechanisms for a SM top quark are
the QCD processes $q \bar q, \ gg \ra t \bar t$ .
For a heavy top quark, $m_t > 100$ GeV, the $q \bar q$ fusion process
becomes most important.
The full next-to-leading-order calculation for these QCD processes was
completed a couple of years ago\refmark{\nloqcd}.
Therefore, the production rates for top quark pairs
at hadron colliders are  well predicted.
If the top quark is as heavy as 140 GeV, then another production
mechanism known as the $W$--gluon fusion process becomes
 important\refmark{\wgone,\wgtwo}.
The production mechanism of the latter process involves the electroweak
interaction, therefore it can probe the electroweak sector of
the theory in contrast to the usual QCD production mechanism
which only probes the QCD interaction when counting the
top quark event rates.

\REF\chiral{S. Weinberg, \jourtpol Phys.~Rev. &{166}, 1568 (1968);
S. Coleman, J. Wess and B. Zumino, \jourtpol Phys.~Rev. &{177}, 2239 (1969);
C. Callan, S. Coleman, J. Wess and B. Zumino, \jourtpol Phys.~Rev. &{177}, 2247
(1969);
M. Chanowitz, H. Georgi and M. Golden, \jourtpol Phys.~Rev.~Lett. &{57}, 2344
(1986).}
\REF\pxi{R.D.~Peccei and
X.~Zhang, \jourtpol Nucl.~Phys. &{B337}, 269 (1990).}
\REF\wgthree{ D. Carlson and C.--P. Yuan, in preparation.}

For heavy top quark production, the dominant subprocess is
$W_L b \ra t$, where $W_L$ is a longitudinally polarized $W$--boson.
Based on the argument of the electroweak chiral lagrangian\refmark{\chiral},
it is likely that new physics
will show up in the interaction of a longitudinal $W$, which is equivalent to
a Goldstone boson characterizing the symmetry breaking mechanism, and a heavy
fermion. Hence, it is important to measure
these form factors to test for the
possibility of having  {\it nonuniversal} gauge couplings
of~\tbW due to some dynamical symmetry breaking scenario\refmark{\pxi}.
It is thus important that the $W$-gluon fusion process
be studied at the Tevatron because any difference in the~\tbW
vertex will cause the production rates for top quarks to  differ from
the SM predictions.
The strategies for  observing a top quark from this process at
the Tevatron were extensively discussed in Ref.~{\wgone}.
A preliminary study shows that it is possible to study top quarks
produced via this process at the SSC and the LHC if a $b$-vertex detector
is used\refmark{\wgthree}. We find that there are about 20,000
signal events and 700 background events for a 140 GeV top quark produced
via the $W$-gluon fusion process at the SSC, where the branching ratios
for $W^\pm \ra e^\pm$ or $\mu^\pm$ are included.
 This result is obtained by assuming that
it is possible to detect a jet within rapidity 4.0 and with a minimum
transverse momentum of 40 GeV.
This result of large signal-to-background ratio is mainly
due to the characteristic features of the transverse momentum and
rapidity distributions of the spectator quark which emitted the virtual $W$
in the $W$-gluon fusion process.
However, if the rapidity coverage of the SSC/LHC detector is smaller,
then the efficiency of keeping signal events is lower and the
background event rates become larger.

\REF\qcdtpol{
F.~Berends et al., in the report of the Top Physics Working
Group from the {\it Proceedings of the Large Hadron
Collider Workshop}, 4-9 October 1990, Aachen, ed. G.~Jarlskog and D.~Rein,
CERN publication CERN 90-10;
A.~Stange and S.~Willenbrock, Fermilab preprint
FERMILAB-PUB-93-027-T, Feb. 1993.}
\REF\qqttqcd{ C. Kao, G. Ladinsky, and C.--P. Yuan, preprint
MSUHEP 93/04, 1993}

\REF\baryon{N.~Turok and J.~Zadrozny {\journal Phys. Rev.&65 (90) 2331};
M.~Dine, P.~Huet, R. Singleton and L.~Susskind
{\journal Phys. Lett.&B257 (91) 351};
L.~McLerran, M. Shaposhnikov, N.~Turok and
M.~Voloshin{\journal Phys. Lett.&B256 (91) 451};
A.~Cohen, D.~Kaplan and A.~Nelson{\journal Phys. Lett.&B263 (91) 86}.}

Top quarks will have longitudinal
polarization if weak effects are present in their production.
The polarization of the top quark produced from the usual QCD process
after including the electroweak radiative
corrections was discussed in Refs.~{\qcdtpol} and {\qqttqcd}.
In the SM, the heavy top quark produced via the $W$--gluon
fusion process is left--handed polarized,
but it is  unpolarized from the usual
QCD production at the Born level.
If new interactions occur, they may manifest themselves in
an enhancement of the polarization effects in
the production of the top
quark via the $W$--gluon fusion process\refmark{\toppol}.
Furthermore, if CP is violated, the production rate of $t$ from
 $ \bar p p(W^+ g) \ra t \bar b X$
would be different from that of $\bar t$ from
$\bar p p(W^- g) \ra \bar t  b X$.
Therefore, one can detect large CP violation effects
by observing the difference in the production rates of $t$ and $\bar t$.
Large CP violating effects are required to have the cosmological
baryon asymmetry produced at the weak phase transition\refmark{\baryon}.
With the large production rate expected for top quarks at the SSC
and the LHC, it
will be possible to make a detailed study of the interactions of the top
quark including polarization effects.

\REF\nucl{
J.D. Jackson, S.B. Treiman and H.W. Wyld, Jr., \jourtpol
Phys.~Rev. & {106} (1957) 517;
R.B. Curtis and R.R. Lewis, \jourtpol Phys.~Rev. & {107} (1957) 543.}
\REF\gunion{
B. Grzadkowski and J.F. Gunion, preprint UCD-92-7, 1992.}

In the $W$--gluon fusion process the top quark is
almost one hundred percent longitudinally
polarized. This allows us to probe CP violation
in the decay process
$t \ra W^+ b \ra l^+ \nu_l b$.
The most obvious observable for this purpose
is
the expectation value of the time--reversal quantity
$
\vec{\bf\sigma_t} \cdot (\widehat{\bf p}_b \times \widehat{\bf p}_{l})
$
where $\vec{\bf\sigma_t}$ is the polarization vector of $t$, and
$\widehat{\bf p}_b$ ($\widehat{\bf p}_l$) is the unit vector of the
$b$ ($l^+$) momentum in the rest frame of the top quark\refmark{\nucl}.
This was suggested in Ref.~\toppol\ and further studied in Ref.~\gunion.

\REF\peskin{C.R.~Schmidt and M.E.~Peskin {\it Phys. Rev. Lett.} {\bf 69} (1992)
410.}
\REF\imkane{ C. Im, G. Kane and P. Malde, preprint UM-TH-92-27, 1992.}

Furthermore, it was shown in Ref.~\peskin\ that it is possible
to study CP violation effects via the usual QCD processes
by observing the asymmetry in the energies
of the two leptons from the decay of the $t \bar t$ pairs.
A detailed study on how to measure the CP violation effects at the SSC and LHC
colliders were given by C. Im and G. Kane in Ref.~\imkane.

\REF\steve{ S. Mrenna and C.--P. Yuan, {\it Phys. Rev.} {\bf D46} (1992) 1007.}
\REF\sdc{ SDC technical Design Report, preprint SDC-92-201, 1992.}

In Ref.~\steve, we showed that the intrinsic width of the top quark
can not be measured at the SSC and the LHC
through the usual QCD processes. For instance,
the intrinsic width of a 140 GeV Standard Model top quark is
about 0.6 GeV, and the full width at half maximum of the reconstructed
top quark invariant mass is about 11 GeV after including the detector
resolution
effects by smearing the final state parton momenta. A similar conclusion was
also given from a hadron level analysis
presented in the SDC Technical Design Report which
concluded that reconstructing the top quark
invariant mass gave a width of 9 GeV for a 150 GeV top quark\refmark{\sdc}.
Can the top quark width~$\width$ be measured better than the factor
$11/0.6 \sim 20$ mentioned above? The answer is yes. As
pointed out in Ref.~\steve,
the width $\width$ can be measured by counting the production rate of top
quarks
from the $W$--$b$ fusion process which is {\it equivalent} to the $W$--gluon
fusion process by a proper treatment of the bottom
quark and the $W$ boson as partons inside the
hadron.
 The $W$--boson which interacts with the $b$--quark to produce the top
quark can be treated as an on--shell boson
in the leading log approximation.
 The moral is that even
under the approximations considered,
a factor of 2 uncertainty in the production
rate for this process gives a
factor of 2 uncertainty in the measurement of $\width$.
This is still much better than what can be done
at the SSC and the LHC through the usual QCD processes.
Therefore, this is an extremely important measurement to be made at
the Tevatron because it directly tests the SM coupling of~\tbW.

\REF\kaner{ For a review, see, G. Kane, preprint UM-TH-91-32, 1991.}

After the top quark is found, one can
measure the branching ratio of $t \ra b W^+(\ra l^+\nu)$ by the
ratio of $(2l+\,jets)$ and $(1l+\,jets)$
event rates from $t \bar t$ production.
Then, a model independent measurement of the decay
width~$\width$ can be made by counting the production rate
of $t$ in the $W$--gluon fusion process.
Should the top quark be found to be non-standard and its properties can not
be described by the SM, we then have to look for some
non-standard decay modes of the top quark\refmark{\kaner}.
Nevertheless, it is still important to measure at least one
partial width~$\width$ precisely to discriminate different new physics
predictions.

\REF\gem{ GEM Letter of Intent, preprint GEM TN-92-49, 1991.}
\REF\lhc{Top Physics Working
Group from the {\it Proceedings of the Large Hadron
Collider Workshop}, 4-9 October 1990, Aachen, ed. G.~Jarlskog and D.~Rein,
CERN publication CERN 90-10.}

The mass of the top quark can be measured within about 1.6\%
based on the studies performed in Refs.~{\sdc}, {\gem} and {\lhc}.
The first method is to study the $e \mu$ mode of the $t \bar t$ pair.
After measuring the branching ratio of $t \ra b W^+(\ra l^+\nu)$,
one can deduce the mass of the top quark from the calculated event rate
predicted by the SM. The second method is to measure the invariant mass
of the top quark via $t \ra b W^+(\ra q_1 \bar q_2)$. The third method, which
gives the best measurement of $m_t$, is to measure the invariant mass
$M_{e\mu}$ of
$e^+$ and $\mu^-$ from the decay of $t \ra b (\ra \mu^-) W^+(\ra e^+\nu) $.
In Refs.~\qqttqcd\ and \steve\ we showed that $M_{e\mu}$ is neither sensitive
to the QCD correction nor to the polarization of the top quark.

\REF\ttH{ W. Marciano and F. Paige,\jourtpol Phys.~Rev.~Lett. &{66} (1991)
2433;
J. Gunion,\jourtpol Phys.~Lett. &{B261} (1991) 510;
R. Kleiss, Z. Kunszt and W. Stirling,
\jourtpol Phys.~Lett. &{B253} (1991) 269.}
\REF\tbH{ J. Diaz-Cruz and O. Sampayo,
\jourtpol Phys.~Lett. &{B276} (1992) 211;
W. Stirling and D. Summers, \jourtpol Phys.~Lett. &{B283} (1992) 411}

\REF\qcdh{S. Dawson, {\it Nucl. Phys.} {\bf B359} (1991) 283;
A. Djouadi, M. Spira and P.~M.~Zerwas, {\it Phys. Lett.} {\bf B264}
(1991) 440.}
\REF\resum{P. Agrawal, J. Qiu and C.--P. Yuan,
in preparation; C.--P. Yuan, {\it Phys. Lett.} {\bf B283} (1992) 395;
R. P. Kauffman, {\it Phys. Rev.} {\bf D45} (1992) 2273.}

In the SM, the top quark gains mass through the Yukawa coupling
to the Higgs boson $H$. To test the dynamics of the generation of the
fermion mass, we should also
measure the coupling of the $Ht \bar t$. The direct measurement of this
coupling can be done by studying the production
of top quarks in $gg, q \bar q \ra t \bar t H$ and
$qg,  {\bar q} g \ra t \bar b H$\refmark{\ttH,\tbH}.
Since the signal rate is not large at the SSC/LHC, it might be difficult to
get a precise measurement from these processes.
If the luminosity of the collider is high enough to produce a
large number of $qg, {\bar q} g \ra t \bar b H$ events, one can
test the relative sign of the $Ht \bar t$ and $H W^+ W^-$ couplings
which depend on the underlying dynamics of symmetry breaking.
Another place to look for the
coupling of $H t \bar t$
is to study the production of the Higgs boson from the gluon fusion
process through a top quark triangle loop. The QCD next-to-leading-order
calculation of this cross section was performed
a couple of years ago\refmark{\qcdh}.
To get a precise measurement on the $H t \bar t$ coupling through the
radiative corrections to the Higgs boson production, we have to know
the kinematics of the Higgs boson after including the multiple gluon
emission effects from the initial state radiation. This calculation of the
QCD-gluon resummation was also performed in Ref.~\resum.

\REF\wwone{
G. Ladinsky and C.--P. Yuan, {\it Phys. Rev.} {\bf D43} (1991) 789;
R. Kauffman and C.--P. Yuan, {\it Phys. Rev.} {\bf D42} 42 (1990) 956.}
\REF\wwtwo{ J. Bagger, V. Barger, K. Cheung, J. Gunion,
T. Han, G. Ladinsky, R. Rosenfeld
and C.--P. Yuan, preprint MSUHEP-93/05, 1993;
 and the references therein.}
\REF\www{
V. Barger, A.L. Stange and R. Phillips, {\it Phys. Rev.} {\bf D45} (1992)
1484.}

Up to now, we only discuss the top quark as a signal.
But, it is also
one of the most important backgrounds in probing new physics
at high energy colliders.
For example, in Refs.~\wwone\ and \wwtwo\
we showed how to suppress the backgrounds
associated with the top quark to study the strongly interacting
longitudinal $W$-boson system in the TeV region to probe the mechanism of the
electroweak symmetry breaking. Top quarks as backgrounds for some other
processes were discussed in Ref.~\www.

In conclusion, we have to study the top quark in detail to test the
SM and to probe new physics. This can be largely done at the Tevatron
and  SSC/LHC colliders.

\ack

I would like to thank P. Agrawal, D. Carlson,
G. Kane, C. Kao, R. Kauffman, G. Ladinsky, S. Mrenna, and  J. Qiu
for collaboration.
This work was supported in part by the Michigan State University
and Texas National Research Laboratory Commission grant RGFY9240.

\refout

\end